# Radio Frequency Identifiers: What are the Possibilities?

Ahmed Elmorshidy, Ph.D.

**Abstract**—this paper defines the components of radio frequency identifiers (RFID). It also explores the different areas and sectors where RFID can be beneficial. The paper discusses the uses and advantages of RFID in deference, consumer packaged goods (CPG), healthcare, logistics, manufacturing, and retail.

——————————— ◆ ———————————

## 1 INTRODUCTION

What is Radio Frequency Identification? Radio-frequency identification (RFID) is the use of an object (typically referred to as an RFID tag) applied to or incorporated into a product, animal, or person for the purpose of identification and tracking using radio waves. Some tags can be read from several meters away and beyond the line of sight of the reader. [16] Most RFID tags contain at least two parts. One is an integrated circuit for storing and processing information, modulating and demodulating a radio-frequency (RF) signal, and other specialized functions. The second is an antenna for receiving and transmitting the signal. There are generally two types of RFID tags: active RFID tags, which contain a battery and can transmit signals autonomously, and passive RFID tags, which have no battery and require an external source to provoke signal transmission.

Today, RFID is used in enterprise supply chain management to improve the efficiency of inventory tracking and management According to the Association for Automatic Identification and Mobility "Radio frequency identification (RFID) is a generic term that is used to describe a system that transmits the identity (in the form of a unique serial number) of an object or person wirelessly, using radio waves." It's grouped under the broad category of automatic identification technologies RFID is in use all around us. If you have ever chipped your pet with an ID tag, used EZPass through a toll booth, or paid for gas using SpeedPass, you've used RFID. In addition, RFID is increasingly used with biometric technologies for security. Unlike ubiquitous UPC bar-code technology, RFID technology does not require contact or line of sight for communication. RFID data can be read through the human body, clothing and non-metallic materials.

## 2 COMPONENTS OF RFID

A basic RFID system consists of three components:

- An antenna or coil
- A transceiver (with decoder)
- A transponder (RF tag) electronically programmed with unique information

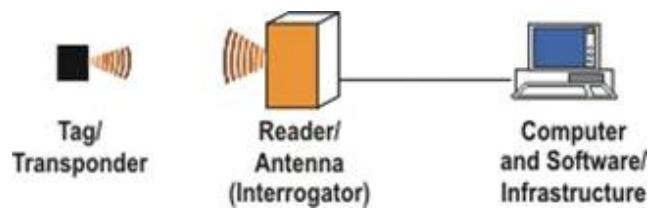

The antenna emits radio signals to activate the tag and to read and write data to it. The reader emits radio waves in ranges of anywhere from one inch to 100 feet or more, depending upon its power output and the radio frequency used. When an RFID tag passes through the electromagnetic zone, it detects the reader's activation signal. The reader decodes the data encoded in the tag's integrated circuit (silicon chip) and the data is passed to the host computer for processing. [1]

————————————————

*Dr. Ahmed Elmorshidy is with Gulf University for Science and Technology, Kuwait. Address: Gulf University for Science & Technology Block 5, Building 1, Mubarak Al-Abdullah Area/West Mishref, Kuwait*
*Dr. Elmorshidy is also affiliated with Claremont Graduate University, California, USA.*

# 3 USES OF RADIO FREQUENCY IDENTIFIERS

## 3.1 DEFENCE

Calling it a turning point for the RFID industry, a group of public and private entities plan this summer to open a business incubator known as the Dayton RFID Convergence Center (DRCC). [incubator known as the Dayton RFID Convergence Center (DRCC). [8] The goal is to bring together entrepreneurs, scientists, technology providers and end users to foster the research, development and marketing of RFID solutions Vikram Sethi, a Dayton RCC board member, as well as the director of Wright State University's (WSU) Institute of Defense Studies and Education (IDSE) and an advisor to its dean, says he first spoke with Proctor about developing an incubator for RFID approximately three years ago. In addition, Sethi says, because the Dayton area includes Wright Patterson Air Force Base—where much of the U.S. military's logistics management is centered—as well as large RFID vendors and customers, an incubator would be centrally located in that city.

Radio frequency identification, along with satellite and cellular technologies, is playing a critical role in the war efforts in Iraq and Afghanistan, according to Major General James L. Hodge, commanding general of the Surface Deployment and Distribution Command (SDDC) division at Scott Air Force Base. [7] At RFID Journal LIVE! 2009, held this week in Orlando, Fla., Hodge explained that RFID enables the U.S. Department of Defense (DOD) to track the movements of cargo to and from these regions.

## 3.2 CONSUMER PACKAGED GOODS (CPG)

Alliance, the merchandising and displays division of Rock-Tenn Co. (one of North America's largest manufacturers of paperboard, containerboard, packaging and merchandising displays), has developed a hardware and service offering that would take promotions management to a new level. Partnering with Seeonic, a provider of inventory-management software, and RFID tag manufacturer UPM Raflatac, Alliance has developed an RFID-enabled system that would allow retailers and producers to track not only new promotional displays, but also the individual items featured on those displays. According to Seeonic, the system could also be used in conjunction with a theft-deterrence system.

Two U.S. grocery-store chains are trialing an electronic shelf label (ESL) system in several of their stores on the East and West coasts. [9] The ESL system, provided by Altierre, comprises RFID-based LCD displays that attach to store shelves to identify products and their prices, enabling immediate shelf-side updates of product data. The system can be managed from a remote location and help the stores avoid using millions of paper labels annually. According to Sunit Saxena, Altierre's chairman and CEO, the system has been six years in the making, and has been piloted for the past nine months. Altierre also opened a retail technology center at its headquarters building in San Jose, Calif. The center, consisting of a supermarket and technology lab, enables potential RFID technology users to test the system in a simulated environment. [9]

## 3.3 HEALTH CARE

Applied Biosystems, a division of biotechnology tool company Life Technologies Corp., is marketing two new lines of genetic analyzers that will employ RFID technology to track the lifespan of consumables as they are used in genetic research, or for diagnostic purposes. The 3500 Series, designed for research use only, will be sold worldwide, to be utilized by universities and laboratories in researching DNA from plants, humans or other animals. The 3500 Dx Series, to be used for medical diagnostic purposes, will be sold only in seven countries in Europe. Either series can also be employed to separate strands of DNA for identification purposes, such as forensics (earlier versions had been used to determine the first map of the human genome). [5]

Reference[5] mentioned that most universities rely on public funding to support their educational and research programs, but the gifts are not always money-based. Each year, for instance the University of California's Anatomical Services department receives nearly a thousand human cadavers, donated to support the education of health professionals, as well as further scientific research. In order to simplify and improve the accuracy of the process of tracking these bodies, the department is currently testing an RFID system developed by the Wireless Internet for Mobile Enterprise Consortium (WINMEC), a UCLA-based research group.

## 3.4 LOGISTICS

Saturdays are busy days at the Federalsburg, Md., cross-docking distribution center of H&M Bay, a national less-than-truckload (LTL) carrier of frozen and refrigerated goods. Late in the day every Friday, the company's fleet of 130 or more trucks begins arriving with trailers loaded with beef, seafood, bakery items and other temperature-sensitive goods. [9] To improve on its system of transferring freight from one truck to another, the company decided in 2008 to install cold-storage rooms so freight could be unloaded and stored in a temperature-controlled fashion until it was ready to be loaded onto another vehicle. However, the firm also realized that if it installed such cold-storage facilities, it would require an RFID-enabled system that would allow it to locate products in those rooms quickly, and to track the length of time each product has been in and out of cold storage, as well as when it shipped.

Now, nine months after launching its new cold-storage operations, the company has concluded that the RFID system has successfully fulfilled those requirements, says John Walker, H&M Bay's software development manager. When Volkswagen drivers in and around Mexico City require replacement parts for their vehicles they rely on a network of VW dealerships. [4] And those dealers rely on a Volkswagen plant in Puebla, located approximately 75 miles outside the city, from which they order these parts. As part of a larger effort initiated by Volkswagen de México in order to improve its spare parts distribution process, cardboard boxes used for shipping spare parts to dealers are being replaced with reusable plastic containers fitted with RFID tags. As part of a larger effort initiated by Volkswagen de México in order to improve its spare parts distribution process, cardboard boxes used for shipping spare parts to dealers are being replaced with reusable plastic containers fitted with RFID tags. The application has enabled the Puebla plant to lower its yearly expenses associated with parts distribution by 30 percent, through reduced packaging costs and an automated shipment-verification process, according to Juan Manuel Rodríguez Flores, sales manager of Grupo Hasar, the RFID systems integrator that deployed the solution for the plant. In late 2007, VW de México enlisted Grupo Hasar to help devise a means by which it could reduce some of the manual steps required to track parts shipments headed to its network of dealers. [4]

### 3.5 MANUFACTURING

RFID Journal (June 1, 2009) and the American Apparel & Footwear Association (AAFA) today announced that Bill Hardgrave, director of the University of Arkansas' RFID Research Center, will present new industry benchmarks related to the benefits that apparel retailers can achieve from using RFID. The presentation will take place at the fourth annual RFID in Fashion conference and exhibition, being held on Aug. 12-13 at the Fashion Institute of Technology in New York City. In April of this year, the RFID Research Center released a research paper indicating that not only does RFID improve inventory accuracy; it can also eliminate the need for annual, manual inventory counts. The paper described the results of the third phase of a pilot studying item-level RFID technology in the retail environment, conducted at three Dillard's stores (see Dillard's, U. of Ark. Study Quantifies RFID's Superiority to Manual Inventory Counts).

AMR Research recently surveyed apparel retailers, asking them to rank the importance of key operational issues they face, and to also rate their ability to respond to such problems. Inventory management and replenishment ranked the highest in terms of importance, though retailers rated their ability to solve the problem as low. (RID Journal June 1, 2009). New Belgium Brewing Co. is using radio frequency identification to track the aluminum-and-steel kegs it utilizes to distribute Fat Tire ale and other beers it produces at its brewery in Fort Collins, Colo. In so doing, the company can track when a particular keg was filled at its facility, shipped to a distributor that delivers it to a bar, restaurant or liquor store, and then returned to the brewery for servicing and refilling. [10]

### 3.6 RETAIL

A Dairy Queen restaurant in the Indianapolis region is providing its customers with RFID-enabled loyalty stickers that attach to their phones, thus enabling them to redeem coupons sent to those handsets via text messages. If the system is successful, Dairy Queen intends to expand it to additional restaurants in the region, as well as nationwide. [14] The system, provided by marketing applications service provider Tetherball, is an expansion of a system Dairy Queen has already been using for 20 months, to send discount coupons for several Indianapolis-area restaurants to customers' cell phones. International Dairy Queen Inc., headquartered in Minneapolis, services more than 5,600 independently owned and operated Dairy Queen stores in the United States, Canada and other countries.

Since late fall 2008 Knox Nursery, located in Winter Garden, Fla., has been using an RFID system that is working so well, its creators at the nursery—together with enterprise software and services company BizSpeed—have begun marketing it to other commercial plant growers and wholesalers. [11] The Grower's Own RFID-based system is commercially available now, while an upgraded version with additional features is slated to become available in the third quarter of 2009. Knox Nursery grows seedlings that it sells to wholesalers and retailers that, in turn, raise the plants until they are large enough for the consumer market. During busy seasons, the company typically sells as many as 30,000 trays of seedlings per week. The nursery employs RFID to identify the disposable plastic trays it sows with seeds—mostly floral annuals, as well as some herbs—then uses the technology to track the inventory and movement of those trays throughout its 13-acre greenhouse, and to expedite the shipping of seedlings to its customers. [11]

## 4 CONCLUSION

Radio-frequency identification (RFID) is the use of an object (typically referred to as an RFID tag) applied to or incorporated into a product, animal, or person for the purpose of identification and tracking using radio waves. Some tags can be read from several meters away and beyond the line of sight of the reader.A basic RFID system consists of three components: An antenna or coil, a transceiver (with decoder), and a transponder (RF tag) electronically programmed with unique information . Radio-frequency identifiers affect many aspects of life and many several types of businesses including defense, retail, consumer good packed goods (CPG), manufacturing, logistics and healthcare. The effect of this new technology is expected to grow and touch several other fields of business especially those types who are technology-intensive.

## RERERENCES

Dr. Ahmed Elmorshidy received his Ph.D. in Management of Information Systems (MIS) in 2004 from Claremont Graduate University, Claremont, California, U.S.A. previously he has earned an MBA in 1995 and an M.A. in Computer Resources and Information Management in 1994 from Webster University, St. Louis, Missouri, U.S.A. Dr. Elmorshidy's B.S. degree was in business administration from Alexandria University, Egypt. Dr. Elmorshidy taught at several academic institutions including Alexandria University, Webster University, Claremont Graduate University, National University, and currently at Gulf University for Science and Technology in Kuwait. Dr. Elmorshidy's research interests are focused on online information systems and the effect of new and disruptive technologies on the field of MIS (Management of Information Systems). Dr. Elmorshidy is a member of IEEE organization and in the Association of Information Systems (AIS).